\documentclass[preprint,aps]{revtex4}
\usepackage{graphicx}

\begin{document}

\title{Superconductors as quantum transducers and antennas for gravitational and electromagnetic radiation}
\author{Raymond Y. Chiao}
\address{Department of Physics\\
University of California\\
Berkeley, CA 94720-7300\\
(E-mail: chiao@physics.berkeley.edu)} 
\date{July 29, 2002 version of PRD (REVTEX) MS}

\begin{abstract}
Superconductors will be considered as macroscopic quantum gravitational
antennas and transducers, which can directly convert upon reflection a beam
of quadrupolar electromagnetic radiation into gravitational radiation, and
vice versa, and thus serve as practical laboratory sources and receivers of
microwave and other radio-frequency gravitational waves. \ An estimate of
the transducer conversion efficiency on the order of unity comes out of the
Ginzburg-Landau theory for an extreme type II, dissipationless
superconductor with minimal coupling to weak gravitational and
electromagnetic radiation fields, whose frequency is smaller than the BCS
gap frequency, thus satisfying the quantum adiabatic theorem. \ The concept
of ``the impedance of free space for gravitational plane waves'' is
introduced, and leads to a natural impedance-matching process, in which the
two kinds of radiation fields are impedance-matched to each other around a
hundred coherence lengths beneath the surface of the superconductor. \ A
simple, Hertz-like experiment has been performed to test these ideas, and
preliminary results will be reported.
\end{abstract}

\maketitle

\address{Department of Physics\\
University of California\\
Berkeley, CA 94720-7300\\
(E-mail: chiao@physics.berkeley.edu)}

\section{Introduction}

In 1966, DeWitt \cite{DeWitt}\ considered the interaction of a
superconductor with gravitational fields, in particular with the
Lense-Thirring field. \ Starting from the general relativistic Lagrangian
for a single electron with a charge $e$ and a mass $m$, he derived in the
limit of weak gravity and slow particles a nonrelativistic Hamiltonian for a
single electron in the superconductor, which satisfied the minimal-coupling
rule%
\begin{equation}
\mathbf{p\rightarrow p}-e\mathbf{A}-m\mathbf{h},
\end{equation}%
where $\mathbf{p}$ is the canonical momentum, $\mathbf{A}$ is the usual
vector potential, and $\mathbf{h}$ is a gauge-like vector potential formed
from the three space-time components $g_{i0}$ of the metric tensor viewed as
an ordinary three-vector. \ Papini \cite{Papini} in 1967 considered the
possibility of the detection the quantum phase shift induced by $\mathbf{h}$
arising from the Lense-Thirring field generated by a nearby rotating massive
body, by means of a superconducting interference device (or SQUID) using
Josephson junctions.\ \ In 1983, Ross \cite{Ross} derived the modified
London equations for a superconductor in a gravitational field, and showed
that these equations are consistent with the modified fluxoid quantization
condition in a gravitational field found earlier by DeWitt in 1966.

In a series of papers in the early 1980s, Anandan and I considered the
possibility of constructing antennas for time-varying Lense-Thirring fields,
and thus for gravitational radiation, using Josephson junctions as
transducers, in $neutral$ superfluid helium analogs of the SQUID using an
antenna geometry in the form of a figure 8 superfluid loop, and also an
antenna bent into a the form of a baseball seam \cite{AnandanChiao}.\ \ In
1985, Anandan \cite{anandan1985}\ considered the possibility of using
superconducting circuits as detectors for astrophysical sources of
gravitational radiation, but did not mention the possibility of
superconductors being efficient emitters, and thus practical laboratory
sources of gravity waves, as is being considered here. \ In 1990, Peng and
Torr used the generalized London equations to treat the interaction of a
bulk superconductor with gravitational radiation, and concluded that such a
superconducting antenna would be many orders of magnitude more sensitive
than a Weber bar \cite{PengTorr}. \ There have also been earlier predictions
of a modified Meissner effect in the response of superconductors to
time-varying Lense-Thirring fields, and hence to gravitational radiation %
\cite{Li1991}\cite{Li1992}. \ For recent work along these lines, see \cite%
{Fischer}. \ These papers, however, also did not consider the possibility of
a transducer action between EM and GR radiation mediated by the
superconductor, as is being considered here. \ Also, the theoretical
approach\ taken here is quite different, as our approach will be based on
the Ginzburg-Landau theory of superconductivity, and the resulting
constitutive relation for the gravitomagnetic field, rather than on the
modified London equations.\ 

Here, I shall show that Josephson junctions, which are difficult to
implement experimentally, are unnecessary, and that a superconductor can by
itself be a $direct$ transducer from electromagnetic to gravitational
radiation upon reflection of the wave from a vacuum-superconductor
interface, with a surprisingly good conversion efficiency. \ By reciprocity,
this conversion process can be reversed, so that gravitational radiation can
also be converted upon reflection into electromagnetic radiation from the
same interface, with equal efficiency. \ The geometry of a superconducting
slab-shaped antenna proposed here is much simpler than some of the earlier
proposed antenna geometries. \ These developments suggest the possibility of
a simple, Hertz-like experiment, in which the emission and the reception of
gravitational radiation at microwave frequencies can be implemented by means
of a pair of superconductors used as transducers. \ Preliminary results of a
first experiment will be reported here.

\section{Superconductors as antennas for gravitational radiation\ }

A strong motivation for performing the quantum calculation to be given
below, is that it predicts a large, counterintuitive \textit{quantum
rigidity }of a macroscopic wavefunction, such as that in a big piece of
superconductor, when it interacts with externally applied gravitational\
radiation fields. \ Mathematically, this quantum rigidity corresponds to the
statement that the macroscopic wavefunction of the superconductor must
remain \textit{single-valued} at all times during the changes arising from
adiabatic perturbations due to radiation fields. \ This implies that objects
such as superconductors should be much better gravitational-wave antennas
than Weber bars \cite{AnandanChiao}\cite{anandan1985}\cite{PengTorr}. \ 

The rigidity of the macroscopic wavefunction\ of the superconductor
originates from the instantaneous Einstein-Podolsky-Rosen (EPR)
correlations-at-a-distance in the behavior of a Cooper pair of electrons in the
Bardeen-Cooper-Schrieffer (BCS) ground state in distant parts of the
superconductor viewed as a single quantum system, when there exists some
kind of gap, such as the BCS gap, which keeps the entire system
adiabatically in its ground state during perturbations due to radiation. \
Two electrons which are members of a single Cooper pair are in a Bohm
singlet state, and hence are quantum-mechanically entangled with each other,
in the sense that they are in a superposition state of opposite spins $and$
opposite momenta. \ This quantum entanglement gives rise to EPR correlations
at long distance scales within the superconductor. \ The electrons in a
superconductor in its ground BCS state are not only $macroscopically$
entangled, but due to the existence of the BCS gap which separates the BCS
ground state energetically from all excited states, they are also $%
protectively$ entangled, in the sense that this entangled state is protected
by the presence of the BCS gap from decoherence arising from the thermal
environment, provided that the system temperature is kept well below the BCS
transition temperature. \ 

The resulting large quantum rigidity is in contrast to the tiny rigidity of
classical matter, such as that of the normal metals used in Weber bars, in
their response to gravitational radiation. \ The essential difference
between quantum and classical matter is that there can exist macroscopic
quantum interference, and hence macroscopic quantum coherence, throughout
the entire quantum system, which is absent in a classical system.\ \ One
manifestation of the tiny rigidity of classical matter is the fact that the
speed of sound in a Weber bar is typically five orders of magnitude less
than the speed of light. \ In order to transfer energy coherently from a
gravitational wave by classical means, for example, by acoustical modes
inside the bar\ to some local detector, e.g., a piezoelectric crystal glued
to the middle of the bar, the length scale of the Weber bar $L$ is limited
to a distance scale on the order of the speed of sound times the period of the
gravitational wave, i.e., an acoustical wavelength $\lambda _{sound}$, which
is typically five orders of magnitude smaller than the gravitational
radiation wavelength $\lambda $ to be detected. \ This makes the Weber bar,
which is thereby limited in its length to $L\simeq \lambda _{sound}$, much
too short an antenna to couple efficiently to free space. \ 

However, macroscopic quantum objects such as superconductors used as
antennas are not limited by these classical considerations, but can have a
length scale $L$ on the same order as (or even much greater than) the
gravitational radiation wavelength $\lambda $. \ Since the radiation
efficiency of a quadrupole antenna scales as the length of the antenna $L$
to the fourth power when $L<<\lambda $, quantum antennas should be much more
efficient in coupling to free space than classical ones like the Weber bar
by at least a factor of $\left( \lambda /\lambda _{sound}\right) ^{4}$. \
Also, we shall see below that a certain type of superconductor may be a $%
transducer$ for directly converting gravitational waves into electromagnetic
waves, and vice versa; this then dispenses altogether with the necessity of
the use of piezoelectric crystals as transducers.

Weinberg \cite{Weinberg}\ gives a measure of the efficiency of coupling of a
Weber bar antenna of mass $M$, length $L$, and velocity of sound $v_{sound}$%
, in terms of a branching ratio for the emission of gravitational radiation
by the Weber bar, relative to the emission of heat, i.e., the ratio of the $%
rate$ of emission of gravitational radiation $\Gamma _{grav}$ relative to
the $rate$ of the decay of the acoustical oscillations into heat $\Gamma
_{heat}$, which is given by%
\begin{equation}
\eta \equiv \frac{\Gamma _{grav}}{\Gamma _{heat}}=\frac{64GMv_{sound}^{4}}{%
15L^{2}c^{5}\Gamma _{heat}}\simeq 3\times 10^{-34}.  \label{WeinbergEff}
\end{equation}%
The quartic power dependence of the efficiency $\eta$\ on the velocity of
sound $v_{sound}$ arises from the quartic dependence of the coupling
efficiency to free space of a quadrupole antenna upon its length $L$, when $%
L<<\lambda $. \ 

Assuming for the moment that the quantum rigidity of a superconductor allows
us to replace the velocity of sound $v_{sound}$ by the speed of light $c$
(i.e., that the typical size $L$ of a quantum antenna bar can become as
large as the wavelength $\lambda $), we see that superconductors can be more
efficient than Weber bars, based on the $v_{sound}^{4}$ factor alone, by
twenty orders of magnitude, i.e., 
\begin{equation}
\left( \frac{c}{v_{sound}}\right) ^{4}\simeq 10^{20}.
\end{equation}%
Thus, even if it should turn out that superconducting antennas in the final
analysis are still not very efficient $generators$ of gravitational
radiation, they should be much more efficient $receivers$ of this radiation
than Weber bars for detecting astrophysical sources of gravitational
radiation \cite{anandan1985}\cite{PengTorr}. \ However, I shall give
arguments below as to why under certain circumstances involving ``natural
impedance matching'' between quadrupolar EM and GR plane waves upon a
mirror-like reflection at the planar surface of extreme type II,
dissipationless superconductors, the efficiency of such superconductors used
as simultaneous transducers and antennas for gravitational radiation, might
in fact become of the order of unity, so that a gravitational analog of
Hertz's experiment might then become possible.\qquad 

But why should the speed of sound in superconductors, which is not much
different from that in normal metals, not also characterize the rigidity of
a superconducting metal when it is in a superconducting state? \ What is it
about a superconductor in its superconducting state that makes it so
radically different from the same metal when it is in a normal state? \ The
answer lies in the important distinction between $longitudinal$ and $%
transverse$ mechanical excitations of the superconductor. \ Whereas
longitudinal, compressional sound waves propagate in superconductors at
normal sound speeds, transverse excitations, such as those\ induced by a
gravitational plane wave incident normally on a slab of superconductor in
its superconducting state, cannot so propagate. \ 

Suppose that the opposite were true, i.e., that there were $no$ substantial
difference between the response of a superconductor and a normal metal to
gravitational radiation. \ (Let us assume for the moment the complete
absence of any electromagnetic radiation.) \ Then the interaction of the
superconductor with gravitational radiation, either in its normal or in its
superconducting state, will be completely negligible, as is indicated by Eq.
(\ref{WeinbergEff}). \ We would then expect the gravitational wave to
penetrate deeply into the interior of the superconductor (see Figure \ref%
{circulations}).

The motion of a Cooper pair deep inside the superconductor (i.e., deep on
the scale of the London penetration depth $\lambda _{L}$ given by Eq. (\ref%
{LondonDepth})), would then be characterized by a velocity $\mathbf{v}%
_{pair}(t)$ which would not be appreciably different from the velocity of a
normal electron or of a nearby lattice ion (we shall neglect the velocity of
sound in this argument, since $v_{sound}<<c$). \ Hence locally, by the weak
equivalence principle, Cooper pairs, normal electrons, and lattice ions
(i.e., independent of the charges and masses of these particles) would all
undergo free fall together, so that 
\begin{equation}
\mathbf{v}_{pair}(t)=-\mathbf{h}(t),
\end{equation}%
where $\mathbf{v}_{pair}(t)$ is the local velocity of a Cooper pair, and
where $-\mathbf{h}(t)$ is the local velocity of a classical test particle,
whose motion is induced by the presence of the gravitational wave, as seen
by an observer sitting in an inertial frame located at the center of mass of
the superconductor. Then the $curl$ of the velocity field $\mathbf{v}%
_{pair}(t)$\ deep inside the superconductor, as seen by this observer, would
be nonvanishing%
\begin{equation}
\mathbf{\nabla \times v}_{pair}(t)=-\mathbf{\nabla \times h}(t)=-\mathbf{B}%
_{G}(t)\neq 0,
\end{equation}%
since the Lense-Thirring or gravitomagnetic field $\mathbf{B}_{G}(t)$ of
gravitational radiation does not vanish deep in the interior of the
superconductor when gravitational radiation is present (we shall see
presently that the $\mathbf{h}(t)$ field plays the role of a gravitomagnetic
vector potential, just like the vector potential $\mathbf{A}(t)$ in the
electromagnetic case). \ 

However, this leads to a contradiction. \ It is well known that for
adiabatic perturbations (e.g., for gravity waves whose frequencies are
sufficiently far below the BCS gap frequency, so that the entire quantum
system remains $adiabatically$ in its ground state), the superfluid velocity
field $\mathbf{v}_{pair}(t)$ deep in the interior of the superconductor
(i.e., at a depth much greater than the London penetration depth) must
remain $irrotational$ at all times \cite{putterman}, i.e.,%
\begin{equation}
\mathbf{\nabla \times v}_{pair}(t)=0.
\end{equation}%
Otherwise, if this irrotational condition were not satisfied in the presence
of gravitational radiation, the wavefunction would not remain single-valued.
\ Deep inside the superconductor, $\mathbf{v}_{pair}(t)=\frac{\hbar }{m_{2}}%
\mathbf{\nabla }\phi (t)$, where $m_{2}$ is the mass of the Cooper pair. \
Thus the superfluid or Cooper pair velocity is directly proportional to the
spatial gradient of the phase $\phi (t)$ of the Cooper-pair condensate
wavefunction. \ It would then follow from such a\ supposed violation of the
irrotational condition that 
\begin{eqnarray}
\Delta \phi (t) &=&\oint_{C}\mathbf{\nabla }\phi (t)\cdot d\mathbf{l}=\frac{%
m_{2}}{\hbar }\oint_{C}\mathbf{v}_{pair}(t)\cdot d\mathbf{l}  \nonumber \\
&=&\frac{m_{2}}{\hbar }\int \int_{S(C)}\mathbf{\nabla \times v}_{pair}(t)\cdot d%
\mathbf{S=-}\frac{m_{2}}{\hbar }\int \int_{S(C)}\mathbf{B}_{G}(t)\cdot d\mathbf{S%
}\neq 0,  \label{Stokes}
\end{eqnarray}%
where\ $\Delta \phi (t)$ is the phase difference after one round-trip back
to the same point around an arbitrary closed curve $C$ deep inside the
superconductor in its ground state, and $S(C)$ is a surface bounded by $C$
(see Figure \ref{circulations}). \ We shall see that the phase shift $\Delta
\phi (t)$\ is a gauge-invariant quantity (see Eq.(\ref{TwinParadox})), and
that, for small circuits $C$, it is directly proportional to a nonvanishing component of the Riemann
curvature tensor. \ However, that the round-trip phase shift $\Delta \phi (t)
$ is nonvanishing in the ground state wavefunction of any quantum system, is
impossible in QM due to the \textit{single-valuedness} of the wavefunction
in the local inertial frame of the center-of-mass of the system. \ 

There is extensive experimental evidence that the single-valuedness of the
wavefunction in QM is not violated. \ For example, the observed quantization
of orbital angular momentum in atoms and molecules constitutes such evidence
on microscopic length scales. \ Also, the observations of quantization of
the circulation of vortices in both superfluids helium of isotope 3 and
helium of isotope 4, and of the quantization of flux in superconductors,
constitute such evidence on macroscopic length scales. \ As a special case
of the latter when the topological winding number is zero, the Meissner
effect is itself evidence for the validity of the principle of the
single-valuedness of the macroscopic wavefunction. \ 

Therefore, in the presence of a gravity wave, Cooper pairs $cannot$ undergo
free fall, and the $transverse$ excitations of the Cooper pair condensate
must remain rigidly irrotational at all times in the adiabatic limit. \ This
leads to a Meissner-like effect in which the Lense-Thirring field $\mathbf{B}%
_{G}(t)$ is expelled, and would seemingly lead to infinite velocities for
the transverse excitations inside the superconductor. \ (It should be noted
here that superluminal, i.e., faster-than-$c$, infinite, and negative, group
velocities for wave packet excitations in a wide variety of classical and
quantum settings have been predicted and observed \cite{Superluminal}. An
analytic function, e.g., a Gaussian wave packet, contains sufficient
information in its early tail such that a causal medium can, during its
propagation, reconstruct the entire wave packet with a superluminal pulse
advancement, and with little distortion. \ Relativistic causality only forbids
the $front$ velocity, which connects causes to their effects, from exceeding
the speed of light $c$, but does not forbid a wave packet's $group$ velocity
from being superluminal.) \ However, such transverse excitations should be
coupled\ to\ perturbations of the metric of spacetime through the
Maxwell-like equations for the time-varying gravitational fields to be
discussed below, and then the speed of such excitations may turn out to be
governed by the vacuum speed of light $c$. \ \qquad\ \ 

\section{Meissner-like effect in the response of a superconductor to
gravitational radiation}

The calculation of the\ $quantum$ response of large objects, for example, a
big piece of superconductor, to weak gravitational radiation, is based on
the concept of $wavefunction$, or \textit{quantum state}, for example, the
BCS state, and proceeds along completely different lines from the
calculation for the $classical$ response of a Weber bar to this radiation,
which is based on the concept of $geodesic$, or \textit{classical trajectory}
\cite{MTW}. \ When the frequency of the gravitational radiation is\ much
less than the BCS gap, the entire superconductor should evolve in time in
accordance with the quantum adiabatic theorem, and should therefore stay
rigidly, i.e., adiabatically, in its ground state. There results a large,
diamagnetic-like \textit{linear response} of the entire superconductor to
externally applied, time-varying gravitational fields. \ This Meissner-like
effect does not alter the geodesic center-of-mass motion of the
superconductor, but radically alters the internal behavior of its electrons,
which are all radically delocalized due instantaneous EPR-correlations within the
superconductor, even at large distances.

\subsection{Calculation of diamagnetic-like coupling energies: The
interaction Hamiltonian}

Consider a gravitational plane wave propagating along the $z$ axis, which
impinges at normal incidence on a piece of superconductor in the form of a
large circular slab of radius $r_{0}$ and of thickness $d$. \ Let the radius 
$r_{0}$ be much larger than the wavelength $\lambda $ of the plane wave, so
that one can neglect diffraction effects. \ Similarly, let $d$ be much
thicker than $\lambda $. \ For simplicity, let the superconductor be at a
temperature of absolute zero, so that only quantum effects need to be
considered. \ The calculation of the coupling energy of the superconductor
in the simultaneous presence of both electromagnetic and gravitational
fields starts from the Lagrangian for a single particle of rest mass $m$ and
charge $e$ (i.e., an electron, but neglecting its spin)%
\begin{equation}
L=-m(-g_{\mu \nu }\dot{x}^{\mu }\dot{x}^{\nu })^{1/2}+eA_{\mu }\dot{x}^{\mu
},
\end{equation}%
from which a minimal-coupling form of the Hamiltonian\ for an electron in a
superconductor, in the limit of $weak$ gravitational fields and $low$
velocities, has been derived by DeWitt \cite{DeWitt}. \ Here we apply this
minimal-coupling Hamiltonian to $pairs$ of electrons (i.e, Cooper pairs in
spin-zero singlet states),%
\begin{equation}
H=\frac{1}{2m_{2eff}}\left( \mathbf{p}-e_{2}\mathbf{A}-m_{2}\mathbf{h}%
\right) ^{2}\text{ in SI units,}  \label{DeWittHamiltonian}
\end{equation}%
where $m_{2}=2m_{e}$ is the rest mass of a Cooper pair, $m_{2eff}$\ is its
effective mass, $e_{2}=2e\ $is its charge, $\mathbf{p}$\ is its canonical
momentum, $\mathbf{A}$\ is the electromagnetic vector potential, and $%
\mathbf{h}$\ is the gravitomagnetic vector potential, which is the
gravitational analog of $\mathbf{A}$ in the case of weak gravity. \ The
gravitomagnetic vector potential $\mathbf{h}$\ is the three-velocity formed
from the space-time components $h_{i0}$ of the small deviations of the
metric tensor $h_{\mu \nu }=g_{\mu \nu }-\eta _{\mu \nu }$ from flat
spacetime (the metric tensor being given by $g_{\mu \nu }$, and the
Minkowski tensor for flat spacetime being given by $\eta _{\mu \nu }=\mathrm{%
diag}(-1,1,1,1)$). \ Thus we shall define%
\begin{equation}
\left. \mathbf{h}\right| _{i}\equiv h_{i0}c\text{ .}
\end{equation}%
It is convenient for performing this calculation to choose the radiation
gauge for both $\mathbf{A}$\ and\ $\mathbf{h}$, so that%
\begin{equation}
\nabla \cdot \mathbf{A}=\nabla \cdot \mathbf{h}=0\mathbf{,}
\label{RadiationGauge}
\end{equation}%
where the chosen coordinate system is that of an inertial frame which
coincides with the freely-falling center of mass of the superconductor at
the origin (this is $not$ the transverse-traceless gauge choice). \ The
physical meaning of\ $\mathbf{h}$\ is that, apart from a sign, it is the
three-velocity of a local, freely-falling test particle as seen by an
observer in an inertial frame located at the center of mass of the
superconductor. \ In Eq. (\ref{DeWittHamiltonian}), we have neglected for
the moment the interactions of the Cooper pairs with each other.

\begin{figure}[tbp]
\includegraphics{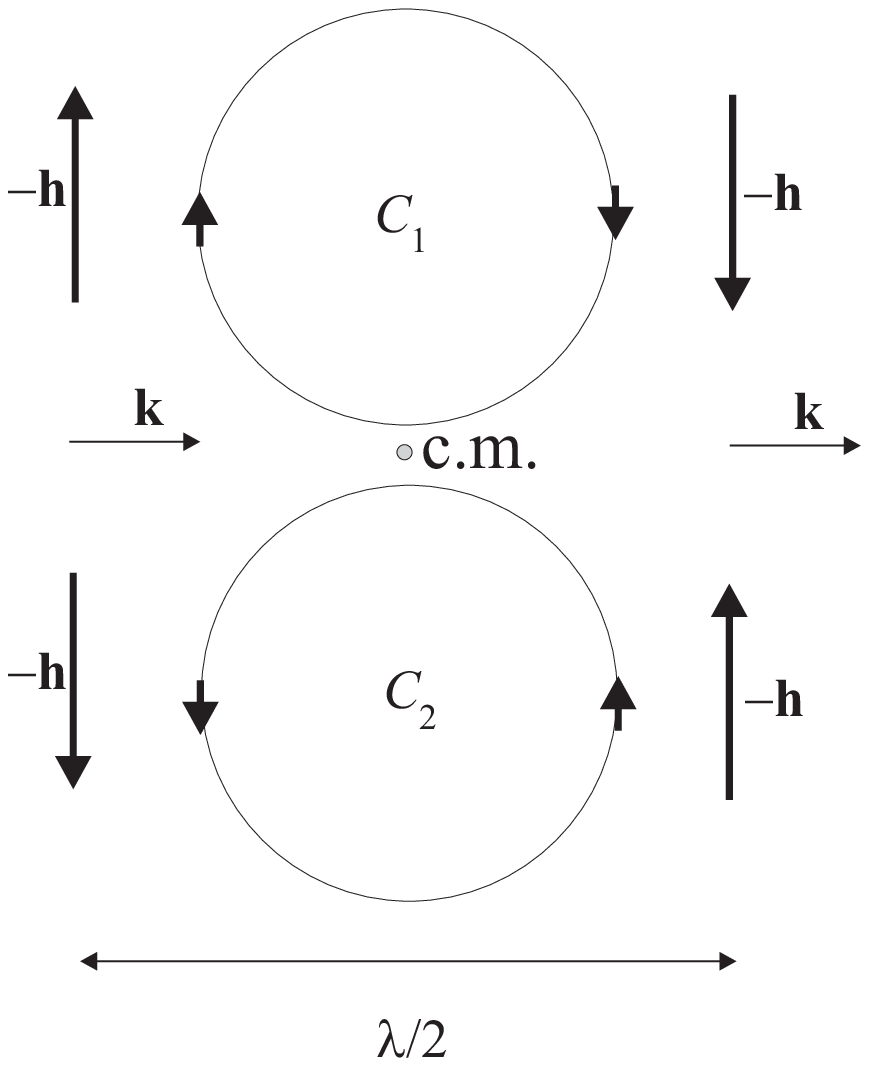}
\caption{A side-view snapshot of a monochromatic gravitational plane wave
inside a thick superconducting slab, propagating to the right with a wave
vector $\mathbf{k}$ along the $z$ axis, which induces tidal motions along
the $x$ axis resulting in the velocity field $\mathbf{-h}$ of test particles,
as seen by an observer sitting in an inertial frame centered on the
center-of-mass (c.m.) of the superconductor. After half a wavelength of
propagation, these tidal motions reverse sign. Hence there exists
nonvanishing circulations around the circuits $C_{1}$ and $C_{2}$, i.e., $%
fluxes$ of the gravitomagnetic (or Lense-Thirring) field inside $C_{1}$ and $%
C_{2}$. These fluxes are directly proportional to the quantities given by
Eqs. (\ref{Stokes}) and (\ref{TwinParadox}), and are gauge-invariant. For small circuits, they
are also directly proportional to the nonvanishing Riemann curvature tensor
component $R_{0x0z}$. The propagation depicted here of a GR wave penetrating
deeply into the interior of a thick slab of superconductor (which is thick
compared to the wavelength $\protect\lambda $) would violate the
single-valuedness of the wavefunction. However, the existence of a
Meissner-like effect, in which all the radiation fields of the GR wave,
including its Lense-Thirring fields, are totally expelled from the
superconductor (apart from a thin surface layer of the order of the London
penetration depth), would prevent such a violation.}
\label{circulations}
\end{figure}

Why not use the standard transverse-traceless gauge in order to perform
these calculations? \ The answer is given in Figure \ref{circulations}, in
which we depict a side view of a snapshot of a gravitational plane wave
propagating to the right. \ The arrows indicate the instantaneous velocity
vectors $-\mathbf{h}$\ of the test particles induced by the wave, as seen by
an inertial observer at the center-of-mass. \ Note that the gravitational
tidal forces reverse in sign after a propagation by half a wavelength to the
right along the $\mathbf{k}$ axis. \ Therefore, by inspection\ of the
diagram, we see that the circulation integrals around circuits $C_{1}$ and $%
C_{2}$ 
\begin{equation}
\oint_{C_{1}}\mathbf{h}\cdot d\mathbf{l}\neq 0,\text{and }\oint_{C_{2}}%
\mathbf{h}\cdot d\mathbf{l}\neq 0  \label{circulation}
\end{equation}%
do not vanish. \ These circulation integrals are gauge-invariant quantities,
since there are related to the gauge-invariant general relativistic time
shift $\Delta t$ (and the corresponding quantum phase shift), where%
\begin{equation}
\Delta t=-\oint_{C}\frac{g_{0i}dx^{i}}{g_{00}}\approx \frac{1}{c}\oint_{C}%
\mathbf{h}\cdot d\mathbf{l}\neq 0.  \label{TwinParadox}
\end{equation}%
For weak gravity, the time shift $\Delta t$ is related through Stokes'
theorem (see Eq. (\ref{Stokes})) to the $flux$ of the gravitomagnetic (or
Lense-Thirring) field through the circuit $C$, which, for small circuits $C$, is directly
proportional to the nonvanishing Riemann curvature tensor component $%
R_{0i0j}$, where the spatial indices $i$ and $j$ are to be contracted with
those corresponding to the small area element enclosed by the circuit $C$
(see Figure \ref{circulations}).  

Since in the transverse-traceless gauge, $\mathbf{h}$ is chosen to be
identically zero, there would be no way to satisfy Eqs. (\ref{circulation})
and (\ref{TwinParadox}). \ In the long-wavelength limit, i.e., in the case
where the antenna is much shorter than a wavelength, such as in Weber bars,
the transverse-traceless gauge can be a valid and more convenient choice
than the radiation gauge being used here. \ However, we wish to\ be able to
consider the case of superconducting slabs which are thick compared with a
wavelength, where the long-wavelength approximation breaks down, and
therefore we cannot use the transverse-traceless gauge, but must use the
radiation gauge instead.

The electromagnetic vector\textbf{\ }potential $\mathbf{A}$\ in the above
minimal-coupling Hamiltonian gives rise to Aharonov-Bohm interference. \ In
like manner, the gravitomagnetic vector potential $\mathbf{h}$ gives rise to
a general relativistic twin paradox for rotating coordinate systems and for
Lense-Thirring fields given by Eq. (\ref{TwinParadox}). \ Therefore $\mathbf{%
h}$ gives rise to Sagnac interference in both light and matter waves. \ The
Sagnac effect has recently been observed in superfluid helium
interferometers using Josephson junctions, and has been used to detect the
Earth's rotation around its polar axis \cite{Simmonds}.

From the above Hamiltonian, we see that the minimal coupling rule for Cooper
pairs now becomes%
\begin{equation}
\mathbf{p\rightarrow p}-e_{2}\mathbf{A}-m_{2}\mathbf{h}\text{ }
\label{Minimal coupling rule}
\end{equation}%
in the simultaneous presence of electromagnetic (EM) and weak general
relativistic (GR) fields. \ This minimal-coupling rule has been
experimentally tested in the case of a uniformly rotating superconducting
ring, since it predicts the existence of a London magnetic moment for the
rotating superconductor, in which magnetic flux is generated through the
center of the ring due to its rotational motion with respect to the local
inertial frame. \ The consequences of the above minimal coupling rule for
the slightly different geometry of a uniformly rotating superconducting
sphere can be easily worked out as follows: Due to the single-valuedness of
the wavefunction, the Aharonov-Bohm and Sagnac phase shifts deep inside the
bulk of the spherical superconductor (i.e., in the interior far away from
the surface) arising from the $\mathbf{A}$ and the $\mathbf{h}$ terms, must
cancel each other exactly. \ Thus the minimal coupling rule leads to a
relationship between the $\mathbf{A}$ and the $\mathbf{h}$ fields inside the
bulk given by 
\begin{equation}
e_{2}\mathbf{A}=-m_{2}\mathbf{h}\text{ .}  \label{A and h}
\end{equation}%
This relationship in turn implies that\ a uniform magnetic field $\mathbf{%
B=\nabla \times A}$, where $\mathbf{A=}\frac{1}{2}\mathbf{B\times r}$ in the
symmetric gauge, will be generated in the interior of the spherical
superconductor due to its uniform rotational motion at an angular velocity $%
\mathbf{\Omega }$ with respect to the local inertial frame, where $\mathbf{%
h=\Omega \times r}$ in the rotating frame. \ \ Thus the London moment effect
will manifest itself here as a uniform magnetic field $\mathbf{B}$ in the
interior of the rigidly rotating sphere, which can be calculated by taking
the curl of both sides of\ Eq. (\ref{A and h}), and yields%
\begin{equation}
\mathbf{B=-}\frac{2m_{2}}{e_{2}}\mathbf{\Omega }\text{ ,}  \label{Larmor}
\end{equation}%
which is consistent with Larmor's theorem. \ In general, the proportionality
constant of the London moment will be given by the inverse of the
charge-to-mass ratio $e_{2}/m_{2}$, where $m_{2}$ has been experimentally
determined to be the $vacuum$ value of the Cooper pair rest mass, apart from
a small discrepancy of the order of ten parts per million, which has not yet
been completely understood \cite{Cabrera}. \ 

However, in the above argument, we have been assuming rigid-body rotation
for the entire body of the superconductor, which is obviously not valid for
microwave-frequency gravitational radiation fields, since the lattice cannot
respond to such high frequencies in such a rigid manner. Hence the above
analysis applies\ only to time-independent (i.e., magnetostatic) and
spatially homogeneous (i.e., uniform) magnetic fields and steady rotations,
and is not valid for the high-frequency, time-dependent, and spatially
inhomogeneous case of the interaction of gravitational and electromagnetic
radiation fields near the surface of the superconductor, since the above
magnetostatic analysis ignores the boundary-value and impedance-matching
problems for radiation fields at the vacuum-superconductor interface, which
will be considered below.

One can generalize the above time-independent minimal-coupling Hamiltonian
to adiabatic time-varying situations as follows:%
\begin{equation}
H=\frac{1}{2m_{2eff}}\left( \mathbf{p}-e_{2}\mathbf{A}(t)-m_{2}\mathbf{h}%
(t)\right) ^{2}\text{ ,}  \label{TimeDependentHamiltonian}
\end{equation}%
where $\mathbf{A}(t)$ and $\mathbf{h}(t)$ are the vector potentials
associated with low-frequency electromagnetic and gravitational radiation
fields, for example. \ (This time-dependent Hamiltonian can also of course
describe low-frequency time-varying tidal and Lense-Thirring fields, as well
as radiation fields, but the adiabatic approximation can still be valid for
radiation fields oscillating at high microwave frequencies, since the BCS
gap frequency of many superconductors lie in the far-infrared part of the
spectrum.) \ Again, it is natural to choose the radiation gauge for both $%
\mathbf{A}(t)$ and $\mathbf{h}(t)$ vector potentials, Eq. (\ref%
{RadiationGauge}), in the description of these time-varying fields. \ The
physical meaning of\ $\mathbf{h}(x,y,z,t)\equiv \mathbf{h}(t)$\ is that it
is the $negative$ of the time-varying three-velocity field $\mathbf{v}%
_{test}(x,y,z,t)$ of a system of noninteracting, locally freely-falling
classical test particles as seen by the observer sitting in an inertial
frame located at the center of mass of the superconductor. \ At first, we
shall treat both $\mathbf{A}(t)$ and $\mathbf{h}(t)$ as classical fields,
but shall treat the matter, i.e., the superconductor, quantum mechanically,
in the standard semiclassical approximation. \ 

The time-dependent Hamiltonian given by Eq. (\ref{TimeDependentHamiltonian})
is, I stress, only a ``guessed'' form of the Hamiltonian, whose ultimate
justification must be an experimental one. \ In case of the time-dependent
vector potential $\mathbf{A}(t)$, there have already been many experiments
which have justified this ``guess,'' but there have been no experiments
which have tested the new term involving $\mathbf{h}(t)$. \ However, one
justification for this new term is that in the static limit, this
``guessed'' Hamiltonian goes over naturally to the magnetostatic
minimal-coupling form, which, as we have seen above, $has$ been tested
experimentally.

From Eq. (\ref{TimeDependentHamiltonian}), we see that the time-dependent
generalization of the minimal-coupling rule for Cooper pairs is%
\begin{equation}
\mathbf{p\rightarrow p}-e_{2}\mathbf{A}(t)-m_{2}\mathbf{h}(t).
\label{t-dependent minimal coupling}
\end{equation}%
It would be hard to believe that one is allowed to generalize $\mathbf{A}$
to $\mathbf{A}(t)$, but that somehow one is $not$ allowed to generalize $%
\mathbf{h}$ to $\mathbf{h}(t)$.

One important consequence that follows immediately from expanding the square
in Eq. (\ref{TimeDependentHamiltonian}) is that there exists a cross-term %
\cite{Solli}%
\begin{equation}
H_{int}=\frac{1}{2m_{2eff}}\left\{ 2e_{2}m_{2}\mathbf{A}(t)\cdot \mathbf{h}%
(t)\right\} =\left( \frac{m_{2}}{m_{2eff}}\right) e_{2}\mathbf{A}(t)\cdot 
\mathbf{h}(t).  \label{Hint}
\end{equation}%
It should be emphasized that Newton's constant $G$ does not enter here. \
The physical meaning of this interaction Hamiltonian $H_{int}$ is that there
should exist a $direct$ coupling between electromagnetic and gravitational
radiation mediated by the superconductor that involves the charge\ $e_{2}$
as its coupling constant, in the quantum limit. \ Thus the strength of this
coupling is electromagnetic, and not gravitational, in its character. \
Furthermore, the $\mathbf{A\cdot h}$\ form of $H_{int}$ implies that there
should exist a \textit{linear and reciprocal} coupling between these two
radiation fields mediated by the superconductor. \ This implies that the
superconductor should be a \textit{quantum-mechanical\ transducer} between
these two forms of radiation, which can, in principle, convert power from
one form of radiation into the other, and vice versa, with equal efficiency.

We can see more clearly the significance of the interaction Hamiltonian $%
H_{int}$ once we convert it into second quantized form and express it in
terms of the creation and annihilation operators for the positive frequency
parts of the\ two radiation fields, as in the theory of quantum optics, so
that in the rotating-wave approximation%
\begin{equation}
H_{int}\propto a^{\dagger }b+b^{\dagger }a
\end{equation}%
where the annihilation operator $a$ and the creation operator $a^{\dagger }$
of a single classical mode of the electromagnetic radiation field, obey the
commutation relation $[a,a^{\dagger }]=1$, and where the annihilation
operator $b$ and the creation operator $b^{\dagger }$ of a matched single
classical mode of the gravitational radiation field, obey the commutation
relation $[b,b^{\dagger }]=1$. \ (This represents a crude, first attempt at
quantizing the gravitational field, which applies only in the case of weak
gravity.) \ The first term $a^{\dagger }b$ then corresponds to the process
in which a graviton is annihilated and a photon is created inside the
superconductor, and similarly the second term $b^{\dagger }a$ corresponds to
the reciprocal process, in which a photon is annihilated and a graviton is
created inside the superconductor. \ Energy is conserved by both of these
processes. \ Time-reversal symmetry, and hence reciprocity, is also
respected by this interaction Hamiltonian.

\subsection{Calculation of diamagnetic-like coupling energies: The
macroscopic wavefunction}

At this point, we need to introduce the purely quantum concept of\textit{\
wavefunction}, in conjunction with the quantum adiabatic theorem. \ To
obtain the response of the superconductor, we must make explicit use of the
fact that the ground state wavefunction of the system is unchanged (i.e.,
``rigid'') during the time variations of\ both $\mathbf{A}(t)$ and $\mathbf{h%
}(t)$. \ The condition for validity of the quantum adiabatic theorem here is
that the frequency of the perturbations $\mathbf{A}(t)$ and $\mathbf{h}(t)$
must be low enough compared with the BCS gap frequency of the
superconductor, so that no transitions are permitted out of the BCS ground
state of the system into any of the excited states of the system. \ However,
``low enough'' can, in practice, still mean quite high frequencies, e.g.,
microwave frequencies in the case of high $T_{c}$ superconductors, so that
it becomes practical for the superconductor to become comparable in size to
the microwave wavelength $\lambda $.

Using the quantum adiabatic theorem, one obtains in first-order perturbation
theory the coupling energy $\Delta E_{int}^{(1)}$ of the superconductor in
the simultaneous presence of both $\mathbf{A}(t)$ and $\mathbf{h}(t)$
fields, which is given by%
\[
\Delta E_{int}^{(1)}=\left( \frac{m_{2}}{m_{2eff}}\right) \left\langle \psi
\left| e_{2}\mathbf{A}(t)\cdot \mathbf{h}(t)\right| \psi \right\rangle =
\]%
\begin{equation}
\left( \frac{m_{2}}{m_{2eff}}\right) \int_{V}dxdydz\text{ }\psi ^{\ast
}(x,y,z)\mathbf{A}(x,y,z,t)\cdot \mathbf{h}(x,y,z,t)\psi (x,y,z)
\end{equation}%
where%
\begin{equation}
\psi (x,y,z)=\left( N/\pi r_{0}^{2}d\right) ^{1/2}=\text{Constant}
\end{equation}%
is the Cooper-pair condensate wavefunction (or Ginzburg-Landau order
parameter) of a homogeneous superconductor of volume $V$ \cite{homogeneous},
the normalization condition having been imposed that%
\begin{equation}
\int_{V}dxdydz\text{ }\psi ^{\ast }(x,y,z)\psi (x,y,z)=N,
\end{equation}%
where $N$ is the total number of Cooper pairs in the superconductor. \
Assuming that both $\mathbf{A}(t)$ and $\mathbf{h}(t)$ have the same (``+'')
polarization of quadrupolar radiation, and that both plane waves impinge on
the slab of superconductor at normal incidence, then in Cartesian
coordinates,%
\begin{equation}
\mathbf{A}(t)=(A_{1}(t),A_{2}(t),A_{3}(t))=\frac{1}{2}(x,-y,0)A_{+}\cos
(kz-\omega t)
\end{equation}%
\begin{equation}
\mathbf{h}(t)=(h_{1}(t),h_{2}(t),h_{3}(t))=\frac{1}{2}(x,-y,0)h_{+}\cos
(kz-\omega t).  \label{h+}
\end{equation}%
One then finds that the time-averaged interaction\ or coupling energy in the
rotating-wave approximation between the electromagnetic and gravitational
radiation fields mediated by the superconductor is%
\begin{equation}
\overline{\Delta E_{int}^{(1)}}=\frac{1}{16}\left( \frac{m_{2}}{m_{2eff}}%
\right) Ne_{2}A_{+}h_{+}r_{0}^{2}.  \label{CouplingEnergy}
\end{equation}%
Note the presence of the factor $N$, which can be very large, since it can
be on the order of Avogadro's number $N_{0}$.

The calculation for the above coupling energy $\overline{\Delta E_{int}^{(1)}%
}$ proceeds along the same lines as that for the Meissner effect of the
superconductor, which is based on the diamagnetism term $H_{dia}$ in the
expansion of the $same$ time-dependent minimal-coupling Hamiltonian, Eq. (%
\ref{TimeDependentHamiltonian}), given by 
\begin{equation}
H_{dia}=\frac{1}{2m_{2eff}}\left\{ e_{2}\mathbf{A}(t)\cdot e_{2}\mathbf{A}%
(t)\right\} .
\end{equation}%
This leads to an energy shift of the system, which, in first-order
perturbation theory, again in the rotating-wave approximation, is given by%
\begin{equation}
\overline{\Delta E_{dia}^{(1)}}=\frac{1}{32m_{2eff}}%
Ne_{2}^{2}A_{+}^{2}r_{0}^{2}.
\end{equation}%
Again, note the presence of the factor $N$, which can be on the order of
Avogadro's number $N_{0}$. \ From this expression, we can obtain the \textit{%
diamagnetic susceptibility} of the superconductor. \ We know from experiment
that the size of this energy shift is sufficiently large to cause a complete
expulsion of the magnetic field from the interior of the superconductor,
i.e., a Meissner effect. \ Hence there must also be a complete reflection of
the electromagnetic wave from the interior of the superconductor, apart from
a thin surface layer of the order of the London penetration depth. \ All
forms of diamagnetism, including the Meissner effect, are purely quantum
effects.

Similarly, there is a ``gravitodiamagnetic'' term $H_{Gdia}$\ in the
expansion of the $same$ minimal-coupling Hamiltonian given by%
\begin{equation}
H_{Gdia}=\frac{1}{2m_{2eff}}\left\{ m_{2}\mathbf{h}(t)\cdot m_{2}\mathbf{h}%
(t)\right\} .
\end{equation}%
This leads to a gravitodiamagnetic energy shift of the system given in
first-order perturbation theory in the rotating-wave approximation by%
\begin{equation}
\overline{\Delta E_{Gdia}^{(1)}}=\frac{1}{32m_{2eff}}%
Nm_{2}^{2}h_{+}^{2}r_{0}^{2}.  \label{GMenergy}
\end{equation}%
From this expression, we can obtain the \textit{gravitodiamagnetic
susceptibility} of the superconductor.

\section{The impedance of free space for gravitational plane waves}

It is not enough merely to calculate the coupling energy arising from the
interaction Hamiltonian given by Eq. (\ref{CouplingEnergy}). \ We must also
compare how large this coupling energy is with respect to the free-field
energies of the uncoupled problem, in particular, that of the gravitational
radiation, in order to see how big an effect we expect to see in the
gravitational sector. \ To this end, I shall introduce the concept of 
\textit{impedance matching}, both between the superconductor and free space
in both forms of radiation, and also between the two kinds of waves inside
the superconductor viewed as a transducer. \ The impedance matching problem
determines the\textit{\ efficiency of power transfer} from the antenna to
free space, and from one kind of wave to the other. \ It is therefore useful
to introduce the concept of the \textit{impedance of free space} $Z_{G}$ for
a gravitational plane wave, which is analogous to the concept of the
impedance of free space $Z_{0}$ for an electromagnetic plane wave (here SI
units are more convenient to use than Gaussian cgs units)%
\begin{equation}
Z_{0}=\frac{E}{H}=\sqrt{\frac{\mu _{0}}{\varepsilon _{0}}}=377\text{ ohms,}
\end{equation}%
where $\mu _{0}$ is the magnetic permeability of free space, and $%
\varepsilon _{0}$ is the dielectric permittivity of free space.

The physical meaning of the ``impedance of free space'' in the
electromagnetic case is that when a plane wave impinges on a large, but
thin, resistive film at normal incidence, due to this film's ohmic losses,
the wave can be substantially absorbed and converted into heat if the
resistance per square element of this film is comparable to 377 ohms. \ In
this case, we say that the electromagnetic plane wave has been approximately
``impedance-matched'' into the film. \ If, however, the resistance of the
thin film is much lower than 377 ohms per square, as is the case for a
superconducting film, then the wave will be reflected by the film. \ In this
case, we say that the wave has been ``shorted out'' by the superconducting
film, and\ that therefore this film reflects electromagnetic radiation like
a mirror. \ By contrast, if the resistance of a normal metallic film is much
larger than 377 ohms per square, then the film is essentially transparent to
the wave. \ As a result, there will be almost perfect transmission. \ 

The boundary value problem for Maxwell's equations coupled to a thin
resistive film with a resistance per square element of $Z_{0}/2$, yields a
unique solution that\ this is the condition for the $maximum$ possible
fractional absorption of the wave energy by the film, which is 50\%, along
with 25\% of the wave energy being transmitted, and the remaining 25\% being
reflected (see Appendix A) \cite{richards}. \ Under such circumstances, we
say that the film has been ``$optimally$ impedance-matched'' to the film. \
This result is valid no matter how thin the ``thin'' film is.

The gravitomagnetic permeability $\mu _{G}$ of free space is \cite{Landau}%
\cite{Forward}%
\begin{equation}
\mu _{G}=\frac{16\pi G}{c^{2}}=3.73\times 10^{-26}\text{ }\frac{\text{m}}{%
\text{kg}}\text{,}  \label{mu_G}
\end{equation}%
i.e., $\mu _{G}$ is the coupling constant which couples the Lense-Thirring
field to sources of mass current density, in the gravitational analog of
Ampere's law for weak gravity. Ciufolini \textit{et al.} have recently
measured, to within $\pm 20\%$, a value of $\mu _{G}$ which agrees with Eq. (%
\ref{mu_G}), in the first observation of the Earth's Lense-Thirring field by
means of laser-ranging measurements of the orbits of two satellites \cite%
{Ciufolini}. From Eq. (\ref{mu_G}), I find that the impedance of free space
is \cite{kraus}%
\begin{equation}
Z_{G}=\frac{E_{G}}{H_{G}}=\sqrt{\frac{\mu _{G}}{\varepsilon _{G}}}=\mu _{G}c=%
\frac{16\pi G}{c}=1.12\times 10^{-17}\text{ }\frac{\text{m}^{2}}{\text{s}%
\cdot \text{kg}}\text{,}  \label{Z_G}
\end{equation}%
where the fact\ has been used that both electromagnetic and gravitational
plane waves propagate at the same speed 
\begin{equation}
c=\frac{1}{\sqrt{\varepsilon _{G}\mu _{G}}}=\frac{1}{\sqrt{\varepsilon
_{0}\mu _{0}}}=3.00\times 10^{8}\text{ }\frac{\text{m}}{\text{s}}\text{.}
\end{equation}%
Therefore, the gravitoelectric permittivity $\varepsilon _{G}$ of free space
is%
\begin{equation}
\varepsilon _{G}=\frac{1}{16\pi G}=2.98\times 10^{8}\text{ }\frac{\text{kg}%
^{2}}{\text{N}\cdot \text{m}^{2}}\text{.}  \label{epsilon_G}
\end{equation}

Newton's constant $G$ now enters explicitly through the expression for the
impedance of free space $Z_{G}$, into the problem of the interaction of
radiation and matter. \ Note that $Z_{G}$ is an extremely small quantity. \
Nevertheless, it is also important to note that it is not strictly zero. \
Since nondissipative quantum fluids, such as superfluids and
superconductors, can in principle have strictly zero losses, they can behave
like ``short circuits'' for gravitational radiation. \ Thus we expect that
quantum fluids, in contrast to classical fluids, can behave like perfect
mirrors for gravitational radiation. \ That $Z_{G}$ is so small explains why
it is so difficult to couple $classical$ matter to gravity waves. \ It is
therefore natural to consider using nondissipative quantum matter instead
for achieving an efficient coupling. \ 

By analogy with the electromagnetic case, the physical meaning of the
``impedance of free space'' $Z_{G}$ is that when a gravitational plane wave
impinges on a large, but thin, $viscous$ fluid film at normal incidence, due
to this film's dissipative losses, the wave can be substantially absorbed
and converted into heat, if the dissipation per square element of this film
is comparable to $Z_{G}$. \ Again in this case, we say that the
gravitational plane wave has been approximately ``impedance-matched'' into
the film. \ If, however, the dissipation of the thin film is much lower than 
$Z_{G}$, as is the case for nondissipative quantum fluids, then the wave
will be reflected by the film. \ In this case, we say that the wave has been
``shorted out'' by the superconducting or superfluid film, and that
therefore the film should reflect gravitational radiation like a mirror. \
By contrast, if the dissipation of the film is much larger than $Z_{G}$, as
is the case for classical matter, then the film is essentially transparent
to the wave, and there will be essentially perfect transmission. \ 

The same boundary value problem holds for the Maxwell-like equations coupled
to a thin viscous fluid film with a dissipation per square element of $%
Z_{G}/2$, and yields the same unique solution that\ this is the condition
for the $maximum$ possible fractional absorption (and the consequent
conversion into heat) of the wave energy by the film, which is 50\%, along
with 25\% of the wave energy being transmitted, and the remaining 25\% being
reflected (see Appendix A). \ Under such circumstances, we again say that
the film has been ``$optimally$ impedance-matched'' to the film. \ Again,
this result is valid no matter how thin the ``thin'' film is.

When the superconductor is viewed as a transducer, the conversion from
electromagnetic to gravitational wave energy, and vice versa, can be viewed
as an $effective$ dissipation mechanism, where instead of being converted
into heat, one form of wave energy is converted into the other form,
whenever impedance matching is achieved within a thin layer inside the
superconductor. \ As we shall see, this can occur naturally when the
electromagnetic wave impedance is $exponentially$ reduced in extreme type II
superconductors as the wave penetrates into the superconductor, so that a
layer is automatically reached in its interior where the electromagnetic
wave impedance is reduced to a level comparable to $Z_{G}$. \ Under such
circumstances, we should expect efficient conversion from one form of wave
energy to the other.

\section{Maxwell-like equations for gravity waves}

For obtaining the impedance of free space $Z_{0}$ for electromagnetic plane
waves, we recall that one starts from Maxwell's equations%
\begin{equation}
\mathbf{\nabla \cdot D}=+\rho _{e}
\end{equation}%
\begin{equation}
\mathbf{\nabla \times E=-}\frac{\partial \mathbf{B}}{\partial t}
\end{equation}%
\begin{equation}
\mathbf{\nabla \cdot B}=0
\end{equation}%
\begin{equation}
\mathbf{\nabla \times H=}+\mathbf{j}_{e}+\frac{\partial \mathbf{D}}{\partial
t},
\end{equation}%
where $\rho _{e}$ is the electrical free charge density (here, the charge
density of Cooper pairs), and $\mathbf{j}_{e}$ is the electrical current
density (due to Cooper pairs), $\mathbf{D}$ is the displacement field, $%
\mathbf{E}$ is the electric field,\ $\mathbf{B}$ is the magnetic induction
field, and $\mathbf{H}$ is the magnetic field intensity. \ The constitutive
relations (assuming an isotropic medium) are%
\begin{equation}
\mathbf{D}=\kappa _{e}\varepsilon _{0}\mathbf{E}
\end{equation}%
\begin{equation}
\mathbf{B}=\kappa _{m}\mu _{0}\mathbf{H}
\end{equation}%
\begin{equation}
\mathbf{j}_{e}=\sigma _{e}\mathbf{E,}
\end{equation}%
where $\kappa _{e}$ is the dielectric constant of the medium, $\kappa _{m}$
is its relative permeability, and $\sigma _{e}$ is its electrical
conductivity. \ We then convert Maxwell's equations into wave equations for
free space in the usual way, and conclude that the speed of electromagnetic
waves in free space is $c=(\varepsilon _{0}\mu _{0})^{-1/2}$, and that the
impedance of free space is $Z_{0}=(\mu _{0}/\varepsilon _{0})^{1/2}$. \ The
impedance-matching problem of a plane wave impinging on a\ thin, resistive
film is solved by using standard boundary conditions in conjunction with the
constitutive relation $\mathbf{j}_{e}=\sigma _{e}\mathbf{E}$.

Similarly, for $weak$ gravity and $slow$ matter, Maxwell-like equations have
been derived from the linearized form of Einstein's field equations \cite%
{Forward}\cite{Braginsky}\cite{Becker}\cite{Tajmar}. \ The gravitoelectric
field $\mathbf{E}_{G}$, which is identical to the local acceleration due to
gravity $\mathbf{g}$,\ is analogous to the electric field $\mathbf{E}$, and
the gravitomagnetic field $\mathbf{B}_{G}$, which is identical to the
Lense-Thirring field, is analogous to the magnetic field $\mathbf{B}$; they
are related to the vector potential $\mathbf{h}$\ in the radiation gauge as
follows:%
\begin{equation}
\mathbf{g=-}\frac{\partial \mathbf{h}}{\partial t}\text{ and }\mathbf{B}_{G}%
\mathbf{=\nabla \times h}\text{ , }  \label{g}
\end{equation}%
which correspond to the electromagnetic relations in the radiation gauge%
\begin{equation}
\mathbf{E=-}\frac{\partial \mathbf{A}}{\partial t}\text{ and }\mathbf{%
B=\nabla \times A}\text{ .}
\end{equation}%
The physical meaning of $\mathbf{g}$ is that it is the three-acceleration of
a local, freely-falling test particle induced by the gravitational
radiation, as seen by an observer in a local inertial frame located at the
center of mass of the superconductor.\ \ The local three-acceleration $%
\mathbf{g}$ is the local time derivative\ of the local three-velocity $-%
\mathbf{h}$ of this test particle, which is a member of a system of
noninteracting, locally freely-falling,\ classical test particles (e.g.,
interstellar dust) with a velocity field $\mathbf{v}_{test}(x,y,z,t)=-%
\mathbf{h}(x,y,z,t)$ as viewed by an observer in the center-of-mass inertial
frame (see Eq. (\ref{g}a)). Similarly, the physical meaning of the
gravitomagnetic field $\mathbf{B}_{G}$ is that it is the local angular
velocity of an inertial frame centered on the same test particle, with
respect to the same observer's inertial frame, which is centered on the
freely-falling center-of-mass of the superconductor. Thus $\mathbf{B}_{G}$\
is the Lense-Thirring field induced by gravitational radiation. \ 

The Maxwell-like equations for weak gravitational fields (upon setting the
PPN (``Parametrized Post-Newton'') parameters to be those of general
relativity) are \cite{Braginsky}%
\begin{equation}
\mathbf{\nabla \cdot D}_{G}=-\rho _{G}
\end{equation}%
\begin{equation}
\mathbf{\nabla \times g=-}\frac{\partial \mathbf{B}_{G}}{\partial t}
\label{Faraday-like}
\end{equation}%
\begin{equation}
\mathbf{\nabla \cdot B}_{G}=0
\end{equation}%
\begin{equation}
\mathbf{\nabla \times H}_{G}\mathbf{=}-\mathbf{j}_{G}+\frac{\partial \mathbf{%
D}_{G}}{\partial t}  \label{Ampere-like}
\end{equation}%
where $\rho _{G}$ is the density of local rest mass in the local rest frame
of the matter, and $\mathbf{j}_{G}$ is the local rest-mass current density
in this frame (in the case of classical matter, $\mathbf{j}_{G}=\rho _{G}%
\mathbf{v}$, where $\mathbf{v}$ is the coordinate three-velocity of the
local rest mass; in the quantum case, see Eq. (\ref{current density})). \
Here $\mathbf{H}_{G}$ is the gravitomagnetic field intensity, and $\mathbf{D}%
_{G}$ is the gravitodisplacement field. \ 

Since the forms of these equations are identical to those of Maxwell's
equations, the same boundary conditions follow from them, and therefore the
same solutions for electromagnetic problems carry over formally to the
gravitational ones. \ These include the solution for\ the optimal
impedance-matching problem for a thin, dissipative film (see Appendix A).

The constitutive relations (assuming an isotropic medium) analogous to those
in Maxwell's theory are%
\begin{equation}
\mathbf{D}_{G}=4\kappa _{GE}\varepsilon _{G}\mathbf{g}  \label{kappa_GE}
\end{equation}%
\begin{equation}
\mathbf{B}_{G}=\kappa _{GM}\mu _{G}\mathbf{H}_{G}  \label{kappa_GM}
\end{equation}%
\begin{equation}
\mathbf{j}_{G}=-\sigma _{G}\mathbf{g}  \label{sigma_G}
\end{equation}%
where $\varepsilon _{G}$ is the gravitoelectric permittivity of free space
given by Eq. (\ref{epsilon_G}), $\mu _{G}$ is the gravitomagnetic
permeability of free space given by Eq. (\ref{mu_G}), $\kappa _{GE}$ is the
gravitoelectric dielectric constant of a medium, $\kappa _{GM}$ is its
gravitomagnetic relative permeability, and $\sigma _{G}$ is the
gravitational analog of the electrical conductivity of the medium, whose
magnitude is inversely proportional to its viscosity. \ It is natural to
choose to define the constitutive relation, Eq. (\ref{sigma_G}), with a
minus sign, so that for $dissipative$ media, $\sigma _{G}$ is always a $%
positive$ quantity. The factor of 4 on the right hand side of Eq. (\ref%
{kappa_GE}) implies that Newton's law of universal gravitation emerges from
Einstein's theory of GR in the correspondence principle limit.

The phenomenological parameters $\kappa _{GE},$ $\kappa _{GM},$ and $\sigma
_{G}$ must be determined by experiment. \ Since there exist no negative
masses which can give rise to a gravitational analog of the polarization of
the medium, we expect that at low frequencies, $\kappa _{GE}\rightarrow 1$.
\ However, because of the possibility of large Meissner-like effects such as
in superconductors, $\kappa _{GM}$ need not approach unity at low
frequencies, but can approach zero instead. \ The gravitodiamagnetic
susceptibility calculated from Eq. (\ref{GMenergy}) should\ lead to a value
of $\kappa _{GM}$ close to zero. \ Also, note that $\kappa _{GM}$ can be
spatially inhomogeneous, such as near the surface of a superconductor. \ 

Again, converting the Maxwell-like equations for weak gravity into a wave
equation for free space in the standard way, we conclude that the speed of
GR waves in free space is $c=(\varepsilon _{G}\mu _{G})^{-1/2}$, which is
identical in GR to the vacuum speed of light, and that the impedance of free
space for GR waves is $Z_{G}=(\mu _{G}/\varepsilon _{G})^{1/2}$, whose
numerical value is given by Eq. (\ref{Z_G}).

It should be stressed here that although the above Maxwell-like equations
look formally identical to Maxwell's, there is an elementary physical
difference between gravity and electricity, which must not be overlooked. \
In electrostatics, the existence of both signs of charges means that both
repulsive and attractive forces are possible, whereas in gravity, only
positive signs of masses, and only attractive gravitational forces between
masses, are observed. \ One consequence of this experimental fact is that
whereas\ it is possible to construct Faraday\ cages that completely screen
out electrical forces, and hence electromagnetic radiation fields, it is
impossible to construct gravitational analogs of such Faraday cages that
screen out gravitoelectric forces, such as Earth's gravity.

However, the gravitomagnetic force can be either repulsive or attractive in
sign, unlike the gravitoelectric force. \ For example, the gravitomagnetic
force between two parallel current-carrying pipes changes sign, when the
direction of the current flow is reversed in one of the pipes, according to
the Ampere-like law Eq. (\ref{Ampere-like}). \ Hence $both$ signs of this
kind of gravitational force are possible. \ One consequence of this is that
gravitomagnetic forces $can$ cancel out, so that, unlike gravitoelectric
fields, gravitomagnetic fields $can$ in principle be screened out of the
interiors of material bodies. \ A dramatic example of this is the complete
screening out of the Lense-Thirring field by superconductors in a
Meissner-like effect, i.e., the complete expulsion of the gravitomagnetic
field from the interior of these bodies, which is predicted by the
Ginzburg-Landau theory given below. \ Therefore the expulsion of
gravitational radiation fields by superconductors can also occur, and\ thus
mirrors for this kind of radiation, although counterintuitive, are not
impossible.

\section{Poynting-like vector and the power flow of gravitational radiation}

In analogy with classical electrodynamics, having obtained the impedance of
free space $Z_{G}$, we are now in a position to calculate the time-averaged
power flow in a gravitational plane through a gravitational analog of
Poynting's theorem in the weak-gravity limit. \ The local time-averaged
intensity of a gravitational plane wave is given by the time-averaged
Poynting-like vector%
\begin{equation}
\overline{\mathbf{S}}=\overline{\mathbf{E}_{G}\mbox{\boldmath $\times$}  \mathbf{H}_{G}}\text{ .}
\end{equation}%
For a plane wave propagating in the vacuum, the local relationship between
the magnitudes of the $\mathbf{E}_{G}$ and $\mathbf{H}_{G}$ fields is given
by%
\begin{equation}
\left| \mathbf{E}_{G}\right| =Z_{G}\left| \mathbf{H}_{G}\right| \text{ .}
\end{equation}%
From Eq.(\ref{g}a), it follows that the local\ time-averaged intensity,
i.e., the power per unit area, of a harmonic plane wave of angular
frequency\ $\omega $ is given by%
\begin{equation}
\left| \overline{\mathbf{S}}\right| =\frac{1}{2Z_{G}}\left| \mathbf{E}%
_{G}\right| ^{2}=\frac{\omega ^{2}}{2Z_{G}}\left| \mathbf{h}\right| ^{2}=%
\frac{c^{3}\omega ^{2}}{32\pi G}\left| h_{0i}\right| ^{2}\text{.}
\end{equation}%
For a Gaussian-Laguerre mode of a quadrupolar gravity-wave beam\ propagating
at 10 GHz with an intensity of a milliwatt per square centimeter, the
velocity amplitude $\left| \mathbf{h}\right| $ is typically%
\begin{equation}
\left| \mathbf{h}\right| \simeq 2\times 10^{-20}\text{ m/s, }
\end{equation}%
or the dimensionless strain parameter $\left| h_{0i}\right| =\left| \mathbf{h%
}\right| /c$ is typically%
\begin{equation}
\left| h_{0i}\right| \simeq 8\times 10^{-31}\text{ ,}
\end{equation}%
which is around ten orders of magnitude smaller than the typical strain
amplitudes observable in the earlier versions of LIGO. \ At first sight, it
would seem extremely difficult to detect such tiny amplitudes. \ However, if
the natural impedance matching process in dissipationless, extreme type II
superconductors to be discussed below can be achieved in practice, then both
the generation and the detection of such small strain amplitudes should not
be impossible.

To give an estimate of the size of the magnetic field amplitudes which
correspond to the above gravitational wave amplitudes, one uses energy
conservation in a situation in which the powers in the EM and GR waves
become comparable to each other in the natural impedance-matching process
described below, where, in the special case of perfect power conversion
(i.e., perfect impedance matching) from EM to GR radiation,%
\begin{equation}
\frac{\omega ^{2}}{2Z_{0}}\left| \mathbf{A}\right| ^{2}=\frac{\omega ^{2}}{%
2Z_{G}}\left| \mathbf{h}\right| ^{2}\text{ ,}
\end{equation}%
from which it follows that 
\begin{equation}
\frac{\left| \mathbf{A}\right| }{\left| \mathbf{h}\right| }=\frac{\left| 
\mathbf{B}\right| }{\left| \mathbf{B}_{G}\right| }=\frac{\left| \mathbf{B}%
\right| }{\left| \mathbf{\Omega }_{G}\right| }=\left( \frac{Z_{0}}{Z_{G}}%
\right) ^{1/2}\text{,}
\end{equation}%
where the ratio is given by the square-root of the impedances of free space $%
(Z_{0}/Z_{G})^{1/2}$ instead of\ the mass-to-charge ratio $2m_{2}/e_{2}$
ratio implied by Eq. (\ref{Larmor}). \ Thus for the\ above numbers%
\begin{equation}
\left| \mathbf{B}\right| \simeq 5\times 10^{-3}\text{ Tesla .}
\end{equation}

\section{Ginzburg-Landau equation coupled to both electromagnetic and
gravitational radiation}

A superconductor in the presence of the electromagnetic field $\mathbf{A}(t)$
alone is well described by the Ginzburg-Landau (G-L) equation for the
complex order parameter $\psi $, which in the adiabatic limit is given by %
\cite{Tinkham} \ 
\begin{equation}
\frac{1}{2m_{2eff}}\left( \frac{\hbar }{i}\mathbf{\nabla }-e_{2}\mathbf{A}%
(t)\right) ^{2}\psi +\beta |\psi |^{2}\psi =-\alpha \psi .  \label{GL}
\end{equation}%
When $\mathbf{A}$ is time-independent, this equation has the same form as
the time-independent Schr\"{o}dinger equation for a particle (i.e., a Cooper
pair) with mass $m_{2eff}$ and a charge $e_{2}$ with an energy eigenvalue $%
-\alpha $, except that there is an extra nonlinear term whose coefficient is
given by the coefficient $\beta $, which arises at a microscopic level from
the Coulomb interaction between Cooper pairs \cite{Tinkham}. \ The values
of these two phenomenological parameters $\alpha $ and $\beta $ must be
determined by experiment. \ There are two important length scales associated
with the two parameters $\alpha $ and $\beta $ of this equation, which can
be obtained by a dimensional analysis of Eq. (\ref{GL}). \ The first is the 
\textit{coherence length}%
\begin{equation}
\xi =\sqrt{\frac{\hbar ^{2}}{2m_{2eff}|\alpha |}}\text{ ,}
\end{equation}%
which is the length scale on which the condensate charge density $e_{2}|\psi
|^{2}$ vanishes, as one approaches the surface of the superconductor from
its interior, and hence the length scale on which the electric field $%
\mathbf{E}(t)$ is screened inside the superconductor. \ The second is the 
\textit{London penetration depth}%
\begin{equation}
\lambda _{L}=\sqrt{\frac{\hbar ^{2}}{2m_{2eff}\beta |\psi |^{2}}}\rightarrow 
\sqrt{\frac{\varepsilon _{0}m_{2eff}c^{2}}{e_{2}^{2}|\psi _{0}|^{2}}}\text{ ,%
}  \label{LondonDepth}
\end{equation}%
which is the length scale on which an externally applied magnetic field $%
\mathbf{B}(t)=\mathbf{\nabla \times A}(t)$ vanishes due to the Meissner
effect, as one penetrates into the interior of the superconductor away from
its surface. Here\ $|\psi _{0}|^{2}$ is the pair condensate density deep
inside the superconductor, where it approaches a constant.

The G-L equation represents a mean field theory of the superconductor at the
macroscopic level, which can be derived from the underlying microscopic BCS
theory \cite{Gorkov}. \ The meaning of the complex order parameter $\psi
(x,y,z)$ is that it is the Cooper pair condensate wavefunction. \ Since $%
\psi (x,y,z)$ is defined as a complex field defined over \textit{ordinary }$%
(x,y,z)$ \textit{space}, it is difficult to discern at this level of
description the underlying quantum entanglement present in the BCS
wavefunction, which is a many-body wavefunction defined over the \textit{%
configuration space} of the many-electron system.\ \ Nevertheless, quantum
entanglement, and hence instantaneous EPR correlations-at-a-distance, shows
up indirectly\ through the nonlinear term $\beta |\psi |^{2}\psi $, and is
ultimately what is responsible for the Meissner effect. \ The G-L theory is
being used here because it is more convenient than the BCS theory for
calculating the response of the superconductor to electromagnetic, and also
to gravitational, radiation.

\begin{figure}[tbp]
\includegraphics{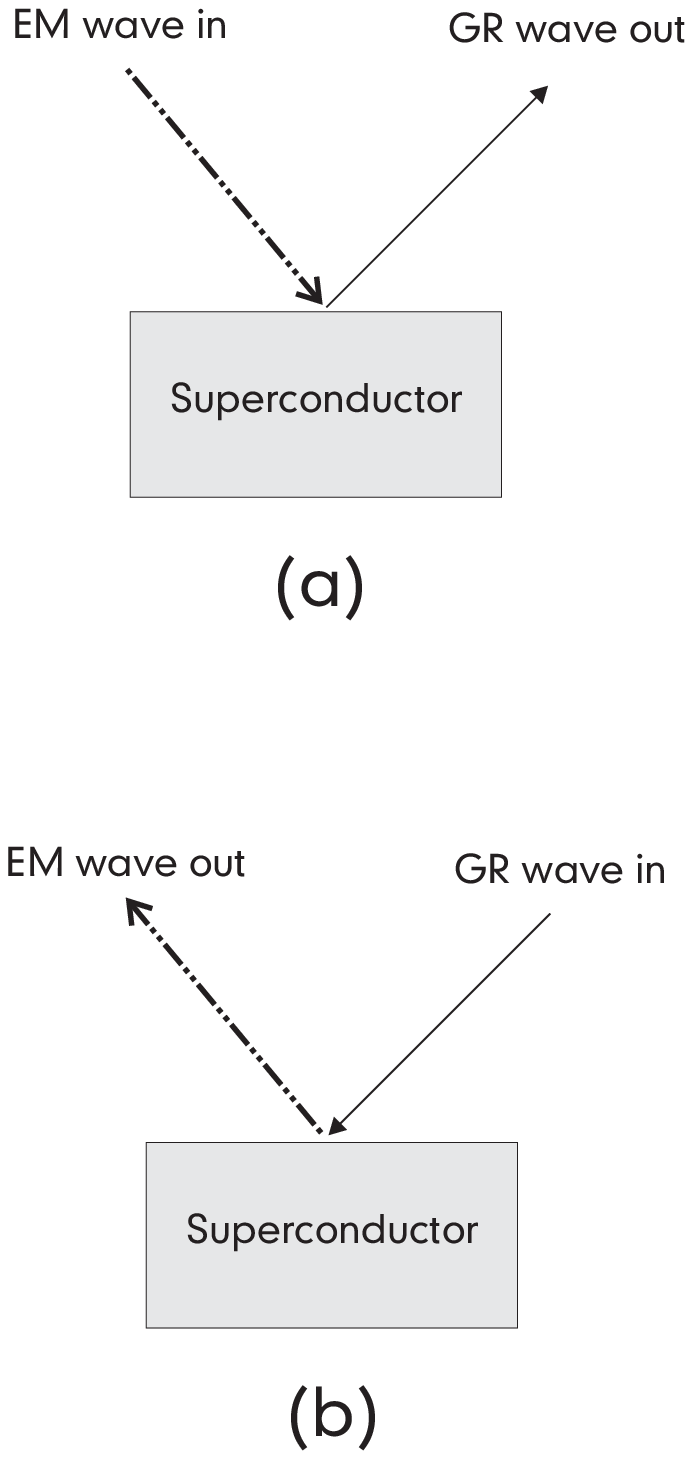}
\caption{Superconductor as\ an impedance-matched transducer between
electromagnetic (EM) and gravitational (GR) radiation. \ (a) An EM plane
wave is converted upon reflection into a GR plane wave. (b) The reciprocal
(or time-reversed) process in which a GR plane wave is converted upon
reflection into an EM plane wave. \ Both EM and GR waves possess the same 
\textit{quadrupolar} polarization pattern.}
\label{Fresnel}
\end{figure}

I would like to propose that the Ginzburg-Landau equation should be
generalized to include gravitational radiation fields $\mathbf{h}(t)$, whose
frequencies lie well below the BCS gap frequency, by using the
minimal-coupling rule, Eq. (\ref{t-dependent minimal coupling}), to the
following equation:%
\begin{equation}
\frac{1}{2m_{2eff}}\left( \frac{\hbar }{i}\mathbf{\nabla }-e_{2}\mathbf{A}%
(t)-m_{2}\mathbf{h}(t)\right) ^{2}\psi +\beta |\psi |^{2}\psi =-\alpha \psi .
\label{Generalized GL equation}
\end{equation}%
Again, the ultimate justification for this equation must come from
experiment. \ With this equation, one can predict what happens at the
interface between the vacuum and the superconductor, when both kinds of
radiation are impinging on this surface at an arbitrary angle of incidence
(see Figure \ref{Fresnel}). \ Since there are still only the two
parameters $\alpha $ and $\beta $ in this equation, there will again be only
the same two length scales $\xi $ and $\lambda _{L}$ that we had {\em before}
adding the gravitational radiation term $\mathbf{h}(t)$. \ Since there are
no other length scales in this problem, one would expect that the
gravitational radiation fields should vanish on the same length scales as
the electromagnetic radiation fields as one penetrates deeply into the
interior of the superconductor. \ Thus one expects there to exist a
Meissner-like expulsion of the gravitational radiation fields from the
interior of the superconductor.

Both $\mathbf{B}(t)$ and $\mathbf{B}_{G}(t)$ fields $must$ vanish into the
interior of the superconductor, since both $\mathbf{A}(t)$ and $\mathbf{h}(t)
$ fields $must$ vanish in the interior. \ Otherwise, the single-valuedness
of $\psi $ would be violated. \ Suppose that $\mathbf{A}(t)$ did not vanish
deep inside the superconducting slab, which is topologically singly
connected. \ Then Yang's nonintegrable phase factor $\exp \left( \left(
ie_{2}\mathbf{/\hbar }\right) \oint \mathbf{A}(t)\cdot d\mathbf{l}\right) $,
which is a gauge-independent quantity,\ would also not vanish \cite{Yang1977}%
, which would lead to a violation of the single-valuedness of $\psi $. \
Similarly, suppose that $\mathbf{h}(t)$ did not vanish. \ Then the
nonintegrable phase factor $\exp \left( \left( im_{2}\mathbf{/\hbar }\right)
\oint \mathbf{h}(t)\cdot d\mathbf{l}\right) $, which is a gauge-independent
quantity, would also\ not vanish, so that again there would be a violation
of the single-valuedness of $\psi $ (see Figure \ref{circulations}).

The $\mathbf{A}(t)$ and $\mathbf{h}(t)$ fields are coupled strongly to each
other through the $e_{2}\mathbf{A}\cdot \mathbf{h}$ interaction Hamiltonian.
\ Since the electromagnetic interaction is very much stronger than the
gravitational one, the exponential decay of $\mathbf{A}(t)$ on the scale of
the London penetration depth $\lambda _{L}$\ should also govern the
exponential decay of the $\mathbf{h}(t)$ field. \ Thus both $\mathbf{A}(t)$
and $\mathbf{h}(t)$ fields decay exponentially with the $same$ length scale $%
\lambda _{L}$ into the interior of the superconductor. \ This implies that
both EM and GR radiation fields will also be expelled from the interior, so
that a flat surface of this superconductor should behave like a plane mirror
for both EM and GR radiation.\ \ 

The Cooper pair current density $\mathbf{j}$, which acts as the source in
Ampere's law in both the Maxwell and the Maxwell-like equations, can be
obtained in a manner similar to that for the Schr\"{o}dinger equation%
\begin{equation}
\mathbf{j}=\frac{\hbar }{2im_{2eff}}\left( \psi ^{\ast }\mathbf{\nabla }\psi
-\psi \mathbf{\nabla }\psi ^{\ast }\right) -\frac{e_{2}}{m_{2eff}}|\psi |^{2}%
\mathbf{A}-\frac{m_{2}}{m_{2eff}}|\psi |^{2}\mathbf{h}\text{ .}
\label{current density}
\end{equation}%
Note thar $\mathbf{j}$ is nonlinear in $\psi $, but linear in $\mathbf{A}$
and $\mathbf{h}$. \ Near the surface of the superconductor, the gradient
terms dominate, but far into the interior, the $\mathbf{A}$ and the $\mathbf{%
h}$ terms dominate. \ We shall now use $\mathbf{j}$ for calculating the
sources for both Maxwell's equations for the electromagnetic fields, and
also for the Maxwell-like equations for the gravitational fields. \ The
electrical current density, the electrical free charge density, the
rest-mass current density, and the rest mass density, are, respectively,%
\begin{equation}
\mathbf{j}_{e}=e_{2}\mathbf{j}\text{ , }\rho _{e}=e_{2}|\psi |^{2}\text{ , }%
\mathbf{j}_{G}=m_{2}\mathbf{j}\text{ , }\rho _{G}=m_{2}|\psi |^{2}\text{ .}
\end{equation}

I have not yet solved the generalized Ginzburg-Landau equation, Eq. (\ref%
{Generalized GL equation}), coupled to both the Maxwell and Maxwell-like
equations through these currents and densities. \ These coupled equations
are nonlinear in $\psi $, but are linear in $\mathbf{A}$ and $\mathbf{h}$
for weak radiation fields. \ However, from dimensional considerations, I can
make the following remarks. \ The electric field $\mathbf{E}(t)$ should be
screened out exponentially towards the interior of the superconductor on a
length scale set by the coherence length $\xi $, since the charge density $%
\rho _{e}=$ $e_{2}|\psi |^{2}$ vanishes exponentially on this length scale
near the surface of the superconductor. \ Similarly, the magnetic field $%
\mathbf{B}(t)$ should vanish exponentially towards the interior of the
superconductor, but on a different length scale set by the London
penetration depth $\lambda _{L} $. \ Both fields vanish exponentially, but
on different length scales. \ 

At first sight, it would seem that similar considerations would apply to the
gravitational fields $\mathbf{g}(t)$ and $\mathbf{B}_{G}(t)$. \ However,
since there exists only one sign of mass for gravity, the gravitoelectric
field $\mathbf{g}(t)$ $cannot$ be screened out. \ Nevertheless, the
gravitomagnetic field $\mathbf{B}_{G}(t)$ $can$ be, indeed $must$ be,
screened by the quantum-mechanical currents $\mathbf{j}$, in order to
preserve the single-valuedness of $\psi $. \ The quantum-current source
terms responsible for the screening out of the $\mathbf{B}_{G}(t)$ field in
Meissner-like effects are not coupled to spacetime by means of Newton's
constant $G$ through the\ right-hand side of the Ampere-like law, Eq. (\ref%
{Ampere-like}), but are coupled directly without the mediation of $G$
through the gravitomagnetic constitutive relation, Eq. (\ref{kappa_GM}), and
through the Faraday-like law, Eq. (\ref{Faraday-like}), whose right-hand
side does not contain $G$. \ For the reasons given above, $\mathbf{B}_{G}(t)$
must decay exponentially on the same scale of length as the gauge field $%
\mathbf{A}(t)$ and the magnetic field $\mathbf{B}(t)$, namely the London
penetration depth $\lambda _{L}$.\ 

The exponential decay into the interior of the superconductor of both EM and
GR waves on the scale of $\lambda _{L}$ means that a flat superconducting
surface should behave like a plane mirror for both electromagnetic and
gravitational radiation. \ However, the behavior of the superconductor as an
efficient $mirror$ is no guarantee that it should also be an efficient $%
transducer$ from one type of radiation to the other. \ For efficient power
conversion, a good transducer impedance-matching process from one kind of
radiation to the other is also required.

\section{Natural impedance matching in extreme type II superconductors \ }

Impedance matching in a natural transduction process between EM and GR waves
could happen near the surface of type II superconductors, where the electric
field decays more quickly than the magnetic field into the interior of the
superconductor, since $\xi <\lambda _{L}$. \ For extreme type II
superconductors, where $\xi <<\lambda _{L}$, the electric field is screened
out much more quickly on the scale of the coherence length $\xi $, than the
magnetic field, which is screened much more slowly on the scale of the
penetration depth $\lambda _{L}$. \ The EM wave impedance for extreme type
II superconductors should therefore decay on the scale of the coherence
length $\xi $ much more quickly than the GR wave impedance, which should
decay much more slowly on the scale of the penetration depth $\lambda _{L}$.
\ The high-temperature superconductor YBCO is an example of an extreme type
II superconductor, for which $\xi $ is less than $\lambda _{L}$ by three
orders of magnitude \cite{Zettl}. \ 

The wave impedance $Z=E/H$ of an EM plane wave depends exponentially as a
function of $z$, the distance from the surface into the interior of the
superconductor, as follows:%
\begin{equation}
Z(z)=\frac{E(z)}{H(z)}=Z_{0}\exp (-z/\xi +z/\lambda _{L}).
\end{equation}%
The GR wave impedance $Z_{G}$, however, behaves very differently, because of
the absence of the screening of the gravitoelectric field, so that $E_{G}(z)$
should be constant independent of $z$ near the surface, and therefore%
\begin{equation}
Z_{G}(z)=\frac{E_{G}(z)}{H_{G}(z)}=Z_{G}\exp (+z/\lambda _{L}).
\end{equation}%
Thus the $z$-dependence of the ratio of the two kinds of impedances should
obey the exponential-decay law 
\begin{equation}
\frac{Z(z)}{Z_{G}(z)}=\frac{Z_{0}}{Z_{G}}\exp (-z/\xi ).
\end{equation}%
We must at this point convert the two impedances $Z_{0}$ and $Z_{G}$ to the
same units for comparison. \ To do so, we express $Z_{0}$ in the natural
units of the quantum of resistance $R_{0}=h/e^{2}$, where $e$ is the
electron charge. \ Likewise, we express $Z_{G}$ in the corresponding natural
units of the quantum of dissipation $R_{G}=h/m^{2}$, where $m$ is the
electron mass. \ Thus we get the $dimensionless$ ratio%
\begin{equation}
\frac{Z(z)/R_{0}}{Z_{G}(z)/R_{G}}=\frac{Z_{0}/R_{0}}{Z_{G}/R_{G}}\exp
(-z/\xi )=\frac{e^{2}/4\pi \varepsilon _{0}}{4Gm^{2}}\exp (-z/\xi ).
\label{zeta}
\end{equation}%
Let us define the ``depth of natural impedance-matching'' $z_{0}$ as the
depth where this dimensionless ratio is unity, and thus natural impedance
matching occurs. \ Then%
\begin{equation}
z_{0}=\xi \ln \left( \frac{e^{2}/4\pi \varepsilon _{0}}{4Gm^{2}}\right)
\approx 97\xi .
\end{equation}%
This result is a robust one, in the sense that the logarithm is very
insensitive to changes in numerical factors of the order of unity in its
argument. \ From this, we conclude that it is necessary penetrate into the
superconductor a distance of\ $z_{0}$, which is around a hundred coherence
lengths $\xi $, for the natural impedance matching process to occur. \ When
this happens, transducer impedance matching occurs automatically, and we
expect that the conversion from electromagnetic to gravitational radiation,
and vice versa, to be an efficient one. \ For example, the London
penetration depth of around 6000 \AA\ in the case of YBCO is very large
compared with $97\xi \simeq $400 \AA\ in this material, so that the
electromagnetic field energy has not yet decayed by much at this natural
impedance-matching plane $z=z_{0}$, although it is mainly magnetic in
character at this point. \ Therefore the transducer power-conversion
efficiency could be of the order of unity, provided that there is no
appreciable parasitic dissipation of the EM radiation fields in the
superconductor before this point.

The Fresnel-like boundary value problem for plane waves incident on the
surface of the superconductor at arbitrary incidence angles and arbitrary
polarizations (see Figure \ref{Fresnel}) needs to be solved in detail
before these conclusions can be confirmed.

\section{Results of a first experiment and future prospects \ }

\begin{figure}[tbp]
\includegraphics{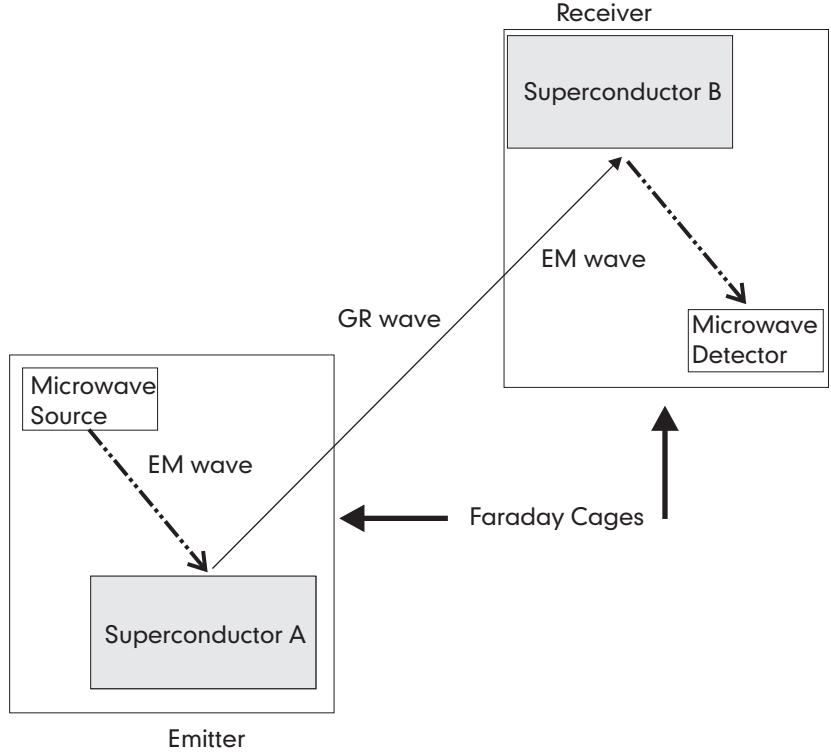}
\caption{Schematic of a simple Hertz-like experiment, in which gravitational
radiation at 12 GHz could be emitted and received using two superconductors.
\ The ``Microwave Source'' generated, by means of a T-shaped quadrupole
antenna, quadrupolar-polarized electromagnetic radiation at 12 GHz (``EM
wave''), which impinged on Superconductor A (a 1 inch diameter, 1/4 inch
thick piece of YBCO), which was placed inside a dielectric Dewar (a stack of
styrofoam cups containing liquid nitrogen), and would be converted upon
reflection into gravitational radiation (``GR wave''). \ The GR wave, but
not the EM wave, could pass through the ``Faraday Cages,'' i.e., normal
metal cans which were lined on the inside with Eccosorb microwave foam
absorbers. \ In the far field of Superconductor A, Superconductor B (also a
1 inch diameter, 1/4 inch thick piece of YBCO in another stack of styrofoam
cups containing liquid nitrogen) would reconvert upon reflection the GR wave
back into an EM wave at 12 GHz, which could then be detected by the
``Microwave Detector,'' which was a sensitive receiver used for microwave
satellite communications, again coupled by another T-shaped quadrupole
antenna to free space. \ The GR wave, and hence the signal at the microwave
detector, should disappear once either superconductor was warmed up above
its transition temperature (90 K), i.e., after the liquid nitrogen boiled
away. }
\label{Hertz}
\end{figure}

However, based on the above crude dimensional and physical arguments, the
prospects for a simple Hertz-like experiment testing these ideas appeared
promising enough that I have performed a first attempt at this experiment
with Walt Fitelson at liquid nitrogen temperature. \ The schematic of this
experiment is shown in Figure \ref{Hertz}. \ Details will be presented
elsewhere.\ 

Unfortunately, we did not detect any observable signal inside the second
Faraday cage, down to a limit of more than 70 dB below the microwave power
source of around $-10$ dBm at 12 GHz. \ (We used a commercial satellite
microwave receiver at 12 GHz with a noise figure of 0.6 dB to make these
measurements; the Faraday cages were good enough to block any direct
electromagnetic coupling by more than 70 dB). \ We checked for the presence
of the Meissner effect in the samples\textit{\ in situ} by observing a
levitation effect upon a permanent magnet by these samples at liquid
nitrogen temperature. \ 

Note, however, that since the transition temperature of YBCO is 90 K, there
may have been a substantial ohmic dissipation of the microwaves due to the
remaining normal electrons at our operating temperature of 77 K, so that the
EM wave was absorbed before it could reach the impedance-matching depth at $%
z_{0}$. \ It may therefore be necessary to cool the superconductor down very
low temperatures before the normal electron component freezes out
sufficiently to achieve such impedance matching. \ The exponential decrease
of the normal electron population at very low tempertures due to the
Boltzmann factor, and thereby an exponential ``freezing out'' of the ohmic
dissipation of the superconductor, may then allow this impedance matching
process to take place, if no other parasitic dissipative processes remain at
these very low temperatures. \ Assuming that the impedance-matching argument
given in Eq. (\ref{zeta}) is correct, and assuming that in the normal state,
the surface resistance of YBCO is on the order of $h/e^{2}=26$ kilohms in
its normal state, one needs a Boltzmann factor of the order of $e^{-100}$ in
order to freeze out the dissipation due to the normal electrons down to an
impedence level comparable to $Z_{G}$. This would imply that temperatures
around a Kelvin should suffice. 

However, there exist unexplained residual microwave and far-infrared losses
in YBCO and other high T$_{\text{c}}$ superconductors, which are independent
of temperature and have a frequency-squared dependence \cite{Miller}. \ It
may therefore be necessary to cool the superconductor down extremely low
temperatures, such as millikelvins, before the normal electron component
freezes out sufficiently to achieve such impedance matching, but more
research needs first to be done to understand the mechanism for these
microwave residual losses before they can be eliminated. An improved
Hertz-like experiment using extreme type II superconductors with extremely
low losses, perhaps at millikelvin temperatures, is a much more difficult,
but worthwhile, experiment to perform.

Such an improved experiment, if successful, would allow us to communicate
through the Earth and its oceans, which, like all classical matter, are
transparent to GR waves. \ Furthermore, it would allow us to directly
observe for the first time the CMB (Cosmic Microwave Background) in GR
radiation, which would tell us much about the very early Universe.

\section{Acknowledgments}

This paper is an abbreviated version of my lecture and book chapter for the
Symposium/Festschrift sponsored by the John Templeton foundation in honor of
Wheeler's 90th birthday. \ I dedicate this paper to my teacher, John
Archibald Wheeler, whose vision helped inspire this work. \ I would like to
thank S. Braunstein, C. Caves, A. Charman, M. L. Cohen, P. C. W. Davies, J.
C. Davis, F. Everitt, W. Fitelson, J. C. Garrison, L. Hall, J. M. Hickmann,
J. M. Leinaas, R. Laughlin, C. McCormick, R. Marrus, M. Mitchell, J. Moore,
J. Myrheim, R. Packard, R. Ramos, P. L. Richards, R. Simmonds, D. Solli, A.
Speliotopoulos, C. H. Townes, W. Unruh, S. Weinreb, E. Wright, W. Wootters,
and A. Zettl for stimulating discussions. \ \ I would especially like to
thank my father-in-law, the late Yi-Fan Chiao, for his financial and moral
support of this work. \ This work was partly supported also by the ONR.

\section{Appendix A: Optimal impedance matching of a gravitational plane
wave into a thin, dissipative film}

Let a gravitational plane wave given by Eq. (\ref{h+}) be normally incident
onto a thin, dissipative (i.e., viscous) fluid film. \ Let the thickness $d$
of this film be arbitrarily thin compared to the gravitational analog of the
skin depth $(2/\kappa _{GM}\mu _{G}\sigma _{G}\omega )^{1/2}$, and to the
wavelength $\lambda $. \ The incident fields calculated using Eqs. (\ref{g}%
)\ (here I shall use the notation $\mathbf{E}_{G}$ instead of $\mathbf{g}$
for the gravitoelectric field) are%
\begin{equation}
\mathbf{E}_{G}^{(i)}=-\frac{1}{2}(x,-y,0)\omega h_{+}\sin (kz-\omega t)
\end{equation}%
\begin{equation}
\mathbf{H}_{G}^{(i)}=-\frac{1}{2Z_{G}}(y,x,0)\omega h_{+}\sin (kz-\omega t).
\end{equation}%
Let $\rho $ be the amplitude reflection coefficient for the gravitoelectric
field; the reflected fields from the film are then%
\begin{equation}
\mathbf{E}_{G}^{(r)}=-\rho \frac{1}{2}(x,-y,0)\omega h_{+}\sin (kz-\omega t)
\end{equation}%
\begin{equation}
\mathbf{H}_{G}^{(r)}=+\rho \frac{1}{2Z_{G}}(y,x,0)\omega h_{+}\sin
(kz-\omega t).
\end{equation}%
Similarly the transmitted fields on the far side of the film are%
\begin{equation}
\mathbf{E}_{G}^{(t)}=-\tau \frac{1}{2}(x,-y,0)\omega h_{+}\sin (kz-\omega t)
\end{equation}%
\begin{equation}
\mathbf{H}_{G}^{(t)}=-\tau \frac{1}{2Z_{G}}(y,x,0)\omega h_{+}\sin
(kz-\omega t),
\end{equation}%
where $\tau $ is the amplitude tranmission coefficient. \ The Faraday-like
law, Eq. (\ref{Faraday-like}), and the Ampere-like law, Eq. (\ref%
{Ampere-like}), when applied to\ the tangential components of the
gravitoelectric and gravitomagnetic fields parallel to two appropriately
chosen infinitesimal rectangular loops which straddle the thin film, lead to
two boundary conditions%
\begin{equation}
\mathbf{E}_{G}^{(i)}+\mathbf{E}_{G}^{(r)}=\mathbf{E}_{G}^{(t)}\text{ and }%
\mathbf{H}_{G}^{(i)}+\mathbf{H}_{G}^{(r)}=\mathbf{H}_{G}^{(t)},
\end{equation}%
which yield the following two algebraic relations:%
\begin{equation}
1+\rho -\tau =0
\end{equation}%
\begin{equation}
1-\rho -\tau =\left( Z_{G}\sigma _{G}d\right) \tau \equiv \zeta \tau 
\end{equation}%
where we have used the constitutive relation, Eq. (\ref{sigma_G}),%
\[
\mathbf{j}_{G}=-\sigma _{G}\mathbf{E}_{G}
\]
to determine the current enclosed by the infinitesimal rectangular loop in
the case of the Ampere-like law, and where we have defined the positive,
dimensionless quantity $\zeta \equiv Z_{G}\sigma _{G}d$. \ The solutions are%
\begin{equation}
\tau =\frac{2}{\zeta +2}\text{ and }\rho =-\frac{\zeta }{\zeta +2}.
\end{equation}%
Using the conservation of energy, we can calculate the absorptivity $A$%
, i.e., the fraction\ of power absorbed from the incident gravitational wave
and converted into heat%
\begin{equation}
A=1-\left| \tau \right| ^{2}-\left| \rho \right| ^{2}=\frac{4\zeta }{(\zeta
+2)^{2}}.
\end{equation}%
To find the condition for maximum absorption, we calculate the derivative $%
dA/d\zeta $ and set it equal to zero. \ The unique solution for maximum
absorptivity occurs at%
\begin{equation}
\zeta =2\text{, where }A=\frac{1}{2}\text{ and }\left| \tau \right| ^{2}=%
\frac{1}{4}\text{ and }\left| \rho \right| ^{2}=\frac{1}{4}\text{.}
\end{equation}%
Thus the optimal impedance-matching condition into the thin, dissipative
film, i.e., when there exists the maximum rate of conversion of
gravitational wave energy into heat, occurs when the dissipation in the
fluid film is $Z_{G}/2$ per square element. \ At this optimum condition,
50\% of the gravitational wave energy will be converted into heat, 25\% will
be transmitted, and 25\% will be reflected. \ This is true independent of
the thickness $d$ of the film, when the film is very thin. \ This solution
is formally identical to that of the optimal impedance-matching problem of
an electromagnetic plane wave into a thin ohmic film \cite{richards}.

\begin{references}
\bibitem{homogeneous} We shall later include the effects of spatial
inhomogeneity near the surface of the superconductor by means of the
Ginzburg-Landau equation, Eq. (\ref{Generalized GL equation}).
\bibitem{richards} I thank my colleague, Paul Richards, for pointing to me
out the solution of the optimal-impedance matching problem for
electromagnetic plane waves, whose extension to gravitational plane waves is
presented in Appendix A.
\bibitem{Zettl} For YBCO (Yttrium Barium Copper Oxide) superconductors, $\xi 
$ is around 3 to 6 \AA\ along the c-axis, and $\lambda _{L}$ along the
c-axis is around 4000 to 8000 \AA . \ I thank my colleague, Alex Zettl, for
providing me with a useful table of the measured superconducting parameters
for YBCO.
\end{references}
\end{document}